\documentclass{jetpl}
\twocolumn

\lat

\title{Effective-field theory  for  high-$T_c$ cuprates}

\rtitle{Effective-field theory  for   high-$T_c$ cuprates}
\sodtitle{Effective-field theory  for   high-$T_c$ cuprates}

\author{A.\,S.\,Moskvin$^{+*}$\/\thanks{e-mail: alexander.moskvin@urfu.ru},
Yu.\,D.\,Panov$^+$}

\rauthor{A.\,S.\,Moskvin, Yu.\,D.\,Panov}

\sodauthor{Moskvin, Panov}

\address{$^+$Ural Federal University, 620083, Ekaterinburg, Russia\\~\\
$^*$Institute of Metal Physics UB RAS, 620108, Ekaterinburg, Russia}

\dates{27 September 2020}{27 September 2020}

\abstract{Starting with a minimal model for the CuO$_2$ planes  with the on-site Hilbert space  reduced to only three effective valence centers [CuO$_4$]$^{7-,6-,5-}$ (nominally Cu$^{1+,2+,3+}$) with different conventional spin and different orbital symmetry we propose a unified non-BCS model that allows one to describe the main features of the phase diagrams of doped cuprates within the framework of a simple effective field theory.

}
\PACS{71.10.Hf, 71.28.+d, 74.20.Mn, 74.72.-h}

\begin{document}

\maketitle





{\bf Introduction}
Today there is no consensus on a theoretical model that allows, within the framework of a single scenario, to describe the phase diagram of the high-$T_c$ cuprates, including  HTSC mechanism itself,  pseudogap phase, strange metal phase, a variety of static and dynamic fluctuations, etc.
In our opinion we miss several fundamental points: inapplicability of the Bardeen-Cooper-Schrieffer (BCS) paradigm with a search for a "superconducting glue"\,,
inapplicability of traditional $\bf k$-momentum (quasi)particle approach, strong but specific electron-lattice effects,
 and inherent intrinsic electronic inhomogeneity in cuprates.
 Recent precision measurements of various physical characteristics on thousands of cuprate samples\,\cite{Bozovic} indicate "insurmountable"\, discrepancies with ideas based on the canonical BCS approach and rather support bosonic mechanism of HTSC cuprates.
A large variety of theoretical models has been designed to account for exotic electronic properties of cuprates and to shed light on their interplay with unconventional superconductivity. However, the most important questions remain unanswered to date.

Earlier we started to develop a  minimal "unparticle"\, model for the CuO$_2$ planes with the "on-site"\, Hilbert space of the CuO$_4$ plaquettes to be a main element of crystal and electron structure of high-T$_c$ cuprates,  reduced to states formed by only three effective valence centers [CuO$_4$]$^{7-,6-,5-}$ (nominally Cu$^{1+,2+,3+}$, respectively), forming a "well isolated"\, charge triplet\,\cite{truegap,dispro,Moskvin-JSNM-2019}. The very possibility of considering these centers on equal footing is predetermined by the strong effects of electron-lattice relaxation in cuprates\,\cite{Mallett,Moskvin-PSS-2020}. The centers are characterized by different conventional spin: s=1/2 for "bare", or "parent" [CuO$_4$]$^{6-}$ center and s=0 for "electron"\, and "hole"\, centers ([CuO$_4$]$^{7-}$- and [CuO$_4$]$^{5-}$-centers, respectively) and different orbital symmetry:$B_{1g}$ for the ground states  of the  [CuO$_4$]$^{6-}$ center,  $A_{1g}$   for the electron center, and the Zhang-Rice $A_{1g}$ or more complicated low-lying non-ZR  states for the hole center.
Electrons of such many-electron atomic species with strong p-d covalence and strong intra-center correlations cannot be described within any conventional (quasi)particle approach that addresses the [CuO$_4$]$^{7-,6-,5-}$    centers within the on-site
hole  representation $|n\rangle$, n = 0, 1, 2, respectively.
Instead of conventional quasiparticle $\bf k$-momentum description we make use of a real space on-site "unparticle"\, S=1 pseudospin formalism to describe the charge triplets and introduce an  effective spin-pseudospin Hamiltonian which takes into account both local and nonlocal correlations, single and two-particle transport, as well as Heisenberg spin exchange interaction. We perform the analysis of the ground state and $T$\,-\,$n$ phase diagrams of the model Hamiltonian by means of a site-dependent variational approach in the grand canonical ensemble within effective field approximation, which treats the on-site quantum fluctuations exactly and all the intersite interactions within the mean-field approximation (MFA) typical for spin-magnetic systems. Within two-sublattice approximation and $nn$-couplings we arrive at several MFA, or N\'{e}el-like phases in CuO$_2$ planes with a single nonzero local order parameter: antiferromagnetic insulator (AFMI), charge order  (CO), glueless $d$-wave Bose superfluid phase (BS), and unusual metallic phase (FL).

{\bf S=1 pseudospin formalism}
To describe the diagonal and off-diagonal, or quantum local charge order we start with a simplified {\it charge triplet model}
that implies a full neglect of spin and orbital degrees of freedom\,\cite{truegap,dispro,Moskvin-JSNM-2019}.
Three charge states of the CuO$_4$ plaquette: a bare center $M^0$=[CuO$_4$]$^{6-}$, a hole center $M^{+}$=[CuO$_4$]$^{5-}$,
and an electron center $M^{-}$=[CuO$_4$]$^{7-}$ are assigned to  three components of the S=1 pseudospin ("isospin") triplet
with the pseudospin projections  $M_S =0,+1,-1$, respectively.

The $S=1$ spin algebra includes the eight independent nontrivial pseudospin operators, the three dipole and five quadrupole operators:
$$
	{\hat S}_z;\, {\hat S}_{\pm}=\mp \frac{1}{\sqrt{2}}({\hat S}_{x}\pm i{\hat S}_{y});\,{\hat S}_z^2;\,{\hat T}_{\pm}=\{{\hat S}_z, {\hat S}_{\pm}\};\,{\hat S}^2_{\pm} \, .
$$
The two Fermi-like  pseudospin raising/lowering operators
${\hat S}_{\pm}$ and ${\hat T}_{\pm}$  change the pseudospin projection by $\pm 1$, with slightly different properties.
In lieu of ${\hat S}_{\pm}$ and ${\hat T}_{\pm}$ operators one may use  two novel operators:
$$
	{\hat P}_{\pm}=\frac{1}{2}({\hat S}_{\pm}+{\hat T}_{\pm});\,{\hat N}_{\pm}=\frac{1}{2}({\hat S}_{\pm}-{\hat T}_{\pm})\,,
$$
which do realize transformations Cu$^{2+}$$\leftrightarrow$\,Cu$^{3+}$ and Cu$^{1+}$$\leftrightarrow$\,Cu$^{2+}$, respectively.
The ${\hat S}_{+}^{2}$ (${\hat S}_{-}^{2}$) operators
create an on-site hole (electron) pair, or composite local boson, with a kinematic constraint $({\hat S}_{\pm}^{2})^2$\,=\,0, that underlines its "hard-core"\, nature.
It should be noted that the effective "quasiparticle"\, wave function of the composite boson has the tetragonal $A_{1g}$-, more precisely, $d_{x^2-y^2}^2$\,-\,symmetry:
\begin{equation}
\Psi ({\bf r})=\Psi (r, \theta , \phi )=R(r)\Theta (\theta)cos^2(2\phi) \, .
\end{equation}
The "on-site"\,, or local S=1 pseudospin state can be written as a superposition
\begin{equation}
	|\Psi\rangle = c_{-1}|1-1\rangle + c_{0}|10\rangle + c_{+1}|1+1\rangle  \, ,
	\label{function1}
\end{equation}
with coefficients that can be represented as follows
\begin{align*}
c_{-1}=\cos\theta\sin\phi e^{-i\alpha};\, c_{0}=\sin\theta e^{i\beta};\,
c_{+1}=\cos\theta\cos\phi e^{i\alpha}	\,.
\end{align*}
The  boson-like  pseudospin raising/lowering operators
${\hat S}_{\pm}^{2}$  define a local "nematic"\, order parameter
\begin{align*}
\langle {\hat S}_{\pm}^{2} \rangle\,=\,\frac{1}{2}\cos^2\theta\sin2\phi\,e^{\mp 2i\alpha} \, .
\end{align*}
 Obviously, this  mean value $\langle {\hat S}_{\pm}^{2} \rangle$ can be addressed to be a complex superconducting local order parameter\,\cite{truegap}.
Unconventional nonzero quantum local mean values of the single-particle operators ${\hat P}_{\pm}$ and ${\hat N}_{\pm}$
\begin{align*}
\langle {\hat P}_{\pm}\rangle =\mp\frac{1}{2}\sin2\theta\cos\phi\,e^{\mp i(\alpha -\beta)}; \\
\langle {\hat N}_{\pm}\rangle =\mp\frac{1}{2}\sin2\theta\sin\phi\,e^{\mp i(\alpha +\beta)}
\end{align*}
imply the local charge density uncertainty.
Strictly speaking, we should extend the on-site Hilbert space to a spin-pseudospin quartet $|SM;s\nu\rangle$: $|1\pm 1;00\rangle$ and $|10;\frac{1}{2}\nu\rangle$, where $\nu =\pm\frac{1}{2}$, and instead of spinless operators ${\hat P}_{\pm}$ and ${\hat N}_{\pm}$ introduce operators ${\hat P}_{\pm}^{\nu}$ and ${\hat N}_{\pm}^{\nu}$, which transform both on-site charge (pseudospin) and spin states as follows
  \begin{align*}
&\hat{P}_{+}^{\nu} | 10; \tfrac{1}{2} \, {-} \nu \rangle = | 11; 00 \rangle ; \quad
\hat{P}_{-}^{\nu} | 11; 00 \rangle = | 10; \tfrac{1}{2} \, \nu \rangle ; \\
&\hat{N}_{+}^{\nu} | 1\, {-}1; 00 \rangle =| 10; \tfrac{1}{2} \, \nu \rangle ; \;\;
\hat{N}_{-}^{\nu} | 10; \tfrac{1}{2} \, {-} \nu \rangle = | 1\, {-}1; 00 \rangle .
\end{align*}
One basic problem  with the local $P_{\pm}^{\nu}$ and $N_{\pm}^{\nu}$ operators and their handling within on-site real-space formalism is their fermionic character.
For the first time the local mean values of fermionic operators similar $\langle {\hat P}_{\pm}^{\nu}\rangle$ and $\langle {\hat N}_{\pm}^{\nu}\rangle$ have been introduced by Caron and Pratt\,\cite{Caron-Pratt} to describe the Hubbard model in the real coordinate space.
At variance with Bose-systems the ground state for kinetic energy in electronic systems is composed, due to the Pauli exclusion principle, of states with different momenta ${\bf k}$ forming the Fermi sea. The problem with local centers is that these only have a limited number of eigenstates and thus seem to be unable to yield any energy bands. However, all of the band states may be easily generated, if to take into account the phase uncertainty of the mean values such as $\langle P_{\pm}^{\nu} \rangle$ and make use of
self-consistency relations reflecting the appropriate Bloch symmetry for the wave vector chosen\,\cite{Doganlar}. As C. Gros\,\cite{Gros} has shown,  the correct ground state energy for non-interacting electrons can be recovered by averaging all of the possible boundary conditions, a method called the "boundary integration technique".

{\bf Effective spin-pseudospin Hamiltonian}
Effective S=1 pseudospin Hamiltonian which does commute with the $z$-component of the total pseudospin  $\sum_{i}S_{iz}$ thus conserving the total charge of the system can be written to be a sum of potential and kinetic energies:
\begin{equation}
{\hat H}={\hat H}_{pot}+{\hat H}_{kin}\, ,
\label{H}	
\end{equation}
where
\begin{equation}
	{\hat H}_{pot} =  \sum_{i}  (\Delta _{i}{\hat S}_{iz}^2
	  - \mu {\hat S}_{iz}) + \sum_{i>j} V_{ij}{\hat S}_{iz}{\hat S}_{jz}\, ,
\label{Hch}	
\end{equation}
with a charge density constraint: $\frac{1}{N}\sum _{i} \langle {\hat S}_{iz}\rangle =n$ ,
where $n$ is the deviation from a half-filling.
The first on-site term in ${\hat H}_{pot}$ describes the effects of a bare pseudospin splitting, or the local energy of $M^{0,\pm}$ centers and relates with the on-site density-density interactions, $\Delta$\,=\,$U$/2, $U$ being the local correlation parameter, or pair binding energy for composite boson. The second term   may be
related to a   pseudo-magnetic field $\parallel$\,$Z$ with $\mu$ being the hole chemical potential.  The third term in ${\hat H}_{pot}$ describes the inter-site density-density interactions, or nonlocal correlations.
Kinetic energy ${\hat H}_{kin}={\hat H}_{kin}^{(1)}+{\hat H}_{kin}^{(2)}$ is a sum of one-particle and two-particle transfer contributions.
In terms of ${\hat P}_{\pm}^{\nu}$ and  ${\hat N}_{\pm}^{\nu}$ operators the  Hamiltonian ${\hat H}_{kin}^{(1)}$ reads as follows:
\begin{equation}
\begin{split}
{\hat H}_{kin}^{(1)}= -\sum_{i>j}\sum_{\nu} [t^p_{ij}{\hat P}_{i+}^{\nu}{\hat P}_{j-}^{\nu}+
 t^n_{ij}{\hat N}_{i+}^{\nu}{\hat N}_{j-}^{\nu}+ \\
 \frac{1}{2} t^{pn}_{ij}({\hat P}_{i+}^{\nu}{\hat N}_{j-}^{\nu}+{\hat P}_{i-}^{\nu}{\hat N}_{j+}^{\nu}) +h.c.] \,.
\end{split}
\label{H1}	
\end{equation}
All the three terms here  suppose a clear physical interpretation. The first $PP$-type term describes one-particle transfer processes:
 Cu$^{3+}$+Cu$^{2+}$$\leftrightarrow$ Cu$^{2+}$+Cu$^{3+}$,
that is  a rather conventional  motion of the hole $M^+$\D centers in the lattice formed by parent $M^0$ (Cu$^{2+}$)\D centers ($p$-type carriers, respectively) or the motion of the $M^0$\D centers in the lattice formed by hole $M^+$\D centers ($n$-type  carriers, respectively).
The second $NN$-type term describes
one-particle transfer processes: Cu$^{1+}$+Cu$^{2+}$$\leftrightarrow$ Cu$^{2+}$+Cu$^{1+}$,
that is  a rather conventional  motion of the electron $M^-$\D centers in the lattice formed by $M^0$\D centers ($n$-type carriers)
or the motion of the $M^0$\D centers in the lattice formed by electron $M^-$\D centers ($p$-type  carriers).
The third $PN$ ($NP$) term in ${\hat H}_{kin}^{(1)}$ defines a very different one-particle transfer process:
Cu$^{2+}$+Cu$^{2+}$$\leftrightarrow$ Cu$^{3+}$+Cu$^{1+}$, Cu$^{1+}$+Cu$^{3+}$,
that is the \emph{local disproportionation/recombination}, or the \emph{electron-hole pair creation/annihilation}. It is this interaction that is responsible for the appearance of carrier sign uncertainty and violation of the "classical"\, Fermi-particle behavior.
Interestingly, the term can be related with a local pairing as the electron $M^-$-center can be addressed to be an electron pair (=\,composite electron boson) localized on the hole $M^+$-center or {\it vice versa} the hole $M^+$-center can be addressed to be a hole pair (=\,composite hole boson) localized on the  electron $M^-$-center.
Hamiltonian ${\hat H}_{kin}^{(2)}$:
\begin{equation}
  {\hat H}_{kin}^{(2)}=-\sum_{i>j} t_{ij}^b({\hat S}_{i+}^{2}{\hat S}_{j-}^{2}+{\hat S}_{i-}^{2}{\hat S}_{j+}^{2})\,,
  \label{H2}
\end{equation}
describes the two-particle (local composite boson) inter-site  transfer, that is the  motion of the electron (hole) center in the lattice formed by the hole (electron) centers, or the exchange reaction:
Cu$^{3+}$+Cu$^{1+}$ $\leftrightarrow$ Cu$^{1+}$+Cu$^{3+}$.
In other words, $t^b_{ij}$ is the transfer integral for the local composite boson.
Depending on the sign of $t^b$, this interaction will stabilize the superconducting $\eta_0$- ($t^b>0$)  or $\eta_{\pi}$- ($t^b<0$) phase.

Conventional Heisenberg spin exchange Cu$^{2+}$\T Cu$^{2+}$ coupling should be transformed  as follows
\begin{equation}
{\hat H}_{ex}=\sum_{i>j} J_{ij}(\hat {\bf s}_i\cdot \hat {\bf s}_j)	\Rightarrow	{\hat H}_{ex}=s^2\sum_{i>j}J_{ij} (\boldsymbol{\sigma}_i\cdot  \boldsymbol{\sigma}_j)\, ,
\end{equation}
where operator $\boldsymbol{\sigma}=2\hat{\rho}^s \mathbf{s}$  takes into account the on-site spin density $\hat{\rho}^s=(1-{\hat S}_{z}^2)$.

Making use of the "Cartesian"\, form of pseudospin operators
\begin{align*}
	 {\hat S}_{\pm}^{2} \,=\,\frac{1}{2}\left(({\hat S}_x^2-{\hat S}_y^2) \pm i\{{\hat S}_x,{\hat S}_y\} \right)={\hat B}_1\pm i{\hat B}_2 \, ; \\
{\hat P}^{\nu}_{\pm}=\frac{1}{2}({\hat P}_1^{\nu}\pm i{\hat P}_2^{\nu}); \,{\hat N}^{\nu}_{\pm}=\frac{1}{2}({\hat N}_1^{\nu}\pm i{\hat N}_2^{\nu}) \nonumber
\end{align*}
with hermitian operators ${\hat B}_{1,2}$, ${\hat P}_{1,2}^{\nu}$, ${\hat N}_{1,2}^{\nu}$ one can
rewrite the spin-pseudospin Hamiltonian in symbolic "vector"\, form as follows
\begin{multline}
	\mathcal{H} =
	\Delta \sum_i {\hat S}_{zi}^2
	+ V \sum_{\left\langle ij\right\rangle} {\hat S}_{zi} {\hat S}_{zj}
	+ Js^2 \sum_{\langle ij \rangle} \boldsymbol{\hat \sigma}_i \boldsymbol{\hat \sigma}_j \\
	- \mathbf{h}s \sum_{i} \boldsymbol{\hat \sigma}_i
	- \mu \sum_i {\hat S}_{zi} 	
	- \frac{t_b}{2} \sum_{\langle ij \rangle} \mathbf{{\hat B}}_{i} \mathbf{{\hat B}}_{j}
	- \frac{t_p}{2} \sum_{\langle ij \rangle \nu} \mathbf{\hat P}_{i}^{\nu} \mathbf{\hat P}_{j}^{\nu} \\
	- \frac{t_n}{2} \sum_{\langle ij \rangle \nu} \mathbf{\hat N}_{i}^{\nu} \mathbf{\hat N}_{j}^{\nu}
	- \frac{t_{pn}}{4} \sum_{\langle ij \rangle \nu}
		\left( \mathbf{\hat P}_{i}^{\nu} \mathbf{\hat N}_{j}^{\nu} + \mathbf{\hat N}_{i}^{\nu} \mathbf{\hat P}_{j}^{\nu} \right) \, ,
\label{HH}
\end{multline}
where we limited ourselves to the interaction of the nearest neighbors, $\boldsymbol{\hat \sigma}= ({\hat \sigma}_x, {\hat \sigma}_y, {\hat \sigma}_z)$, $\mathbf{{\hat B}}=({\hat B}_1, {\hat B}_2)$, $\mathbf{\hat P}^{\nu}=({\hat P}^{\nu}_1, {\hat P}^{\nu}_2)$, $\mathbf{\hat N}^{\nu}=({\hat N}^{\nu}_1, {\hat N}^{\nu}_2)$.

{\bf Effective-field approximation}
Simple effective-field or mean-field theory is as always a good starting point to provide physically clear semi-quantitative description of strongly correlated systems.
Making use of local order parameters without switching to the momentum ${\bf k}$-representation is a typical way to describe "classical"\, phases for spin magnetic systems such as simple N\'eel order.

Hereafter, we perform the analysis of the ground state and $T$\,-\,$n$ phase diagrams of the model Hamiltonian  (\ref{HH}) by means of a site-dependent variational approach (VA) in the grand canonical ensemble within effective-field approximation, which treats the on-site correlation term exactly and all the intersite interactions within the MFA typical for spin-magnetic systems\,\cite{Kapcia}.


We start with assuming the existence of two interpenetrating lattices ($A$ and $B$), restricting the analysis to the two-sublattice solutions for the  single nonzero local order parameter phases. In such a case we introduce 14 parameters of an uniform and 14 parameters of a non-uniform, or staggered order,  as follows:
\begin{equation}
O_{\pm}=\frac{1}{2}(O_A\pm O_B) \, ,
\end{equation}
where $O_{A,B}$ are local order parameters $\langle {\hat S}_{z}\rangle , \langle \boldsymbol{\hat \sigma}\rangle , \langle \mathbf{{\hat B}}\rangle , \langle \mathbf{\hat P}^{\nu}\rangle , \langle \mathbf{\hat N}^{\nu}\rangle $ for $A,B$ sublattice. The corresponding parameters of uniform and staggered order will be denoted below as  $n, \mathbf{m}, \mathbf{B_0}, \mathbf{P}^{\nu}, \mathbf{N}^{\nu}$ and $L_z, \mathbf{l}, \mathbf{B_{\pi}}, \mathbf{P}_L^{\nu}, \mathbf{N}_L^{\nu}$, respectively ($n$\,is a doping level).
The resulting Hamiltonian can be rewritten as a sum of one-site Hamiltonians as follows
\begin{multline}
	\quad \quad \mathcal{\hat H}_0 = \sum_{c=1}^{N/2} \mathcal{\hat H}_c
	,\qquad
	\mathcal{\hat H}_c = \mathcal{\hat H}_A + \mathcal{\hat H}_B \, ,\\
	\mathcal{\hat H}_{\alpha} = \Delta {\hat S}_{z\alpha}^2
	- \left( H_z \pm H_z^L \right) {\hat S}_{z\alpha}
	- \left( \mathbf{h} \pm \mathbf{h}^{l} \right) \boldsymbol{\hat \sigma}_{\alpha}
	- \left( \mathbf{h}_b \pm \mathbf{h}_b^{L} \right) \mathbf{{\hat B}}_{\alpha} \\
		{}- \sum_{\nu} \left( \mathbf{h}_p^{\nu} \pm \mathbf{h}_p^{L,\nu} \right) \mathbf{{\hat P}}_{\alpha}^{\nu}
	- \sum_{\sigma} \left( \mathbf{h}_n^{\nu} \pm \mathbf{h}_n^{L,\nu} \right) \mathbf{{\hat N}}_{\alpha}^{\nu}\, ,
\label{H0}
\end{multline}
where $\alpha=A,B$, the upper (lower) sign corresponds to $A$\,($B$) sublattice,
$H_+$=$H_z$, $\mathbf{h}$, $\mathbf{h}_b$, $\mathbf{h}_p^{\nu}$, $\mathbf{h}_n^{\nu}$, and  $H_-$=$H_z^L$,  $\mathbf{h}^{l}$,  $\mathbf{h}_b^{L}$,
 $\mathbf{h}_p^{L,\nu}$,  $\mathbf{h}_n^{L,\nu}$
($\nu =\,\uparrow,\,\downarrow$)
are uniform and staggered molecular fields, respectively.
Using the partition function
$$
	Z_c = \mathrm{Tr\,} \left( e^{-\beta \mathcal{H}_c } \right)
	= \mathrm{Tr\,} \left( e^{-\beta \mathcal{H}_A } \right) \; \mathrm{Tr\,} \left( e^{-\beta \mathcal{H}_B } \right)
	= Z_A Z_B \, 	,
$$
where $\beta=1/k_BT$, we obtain the expressions for the charge density $n$ and other order parameters as follows:
\begin{equation}
	O_{\pm} = \frac{1}{2\beta} \frac{\partial \ln Z_c}{\partial H_{\pm}}
	,\qquad
	n = \frac{1}{2\beta} \frac{\partial \ln Z_c}{\partial H_z}
	, \, ... .
	\label{eq:PP1}
\end{equation}
The variational approach that will be employed is based on the Bogolyubov inequality for the grand potential $\Omega(\mathcal{H})$:
$$
\Omega(\mathcal{H}) = \Omega(\mathcal{H}_0) + \left\langle \mathcal{H} - \mathcal{H}_0 \right\rangle \, ,
$$
where $\mathcal{H}$ is the Hamiltonian under study (\ref{HH}), $\mathcal{H}_0 $ is the trial
Hamiltonian (\ref{H0}) which depends on the
variational order parameters and can be exactly solved, the thermal average  is
taken over the ensemble defined by $\mathcal{H}_0 $.
We estimate the free energy of the system per one site, $f = \Omega/N + \mu n$, as follows:
\begin{multline}
	\quad \quad f =
	-\frac{1}{2\beta} \ln Z_c
	+ 2V \left( n^2 - L_z^2 \right)
	+{}\\
	{}
	+ 2Js^2 \left( \mathbf{m}^2 - \mathbf{l}^2 \right)
	- t_b \left( \mathbf{B}_0^2 - \mathbf{B}_{\pi}^2 \right)
	-{}\\[0.5em]
	{}	
	- t_p \sum_{\nu} \left( {\mathbf{P}^{\nu}}^2 - {\mathbf{P}_L^{\nu}}^2 \right)
	- t_n \sum_{\nu} \left( {\mathbf{N}^{\nu}}^2 - {\mathbf{N}_L^{\nu}}^2 \right)
	-{}\\
	{}	
	- t_{pn} \sum_{\nu}
		\left( \mathbf{P}^{\nu} \mathbf{N}^{\nu} - \mathbf{P}_L^{\nu} \mathbf{N}_L^{\nu} \right)
	+{}\\
	{}+
	H_z n + H_z^L L_z
	+ \mathbf{h} \mathbf{m}	+ \mathbf{h}^l \mathbf{l}
	+ \mathbf{h}_b \mathbf{B}_0 + \mathbf{h}_b^L \mathbf{B}_{\pi}
	+{}\\[0.5em]
	{}
	+ \sum_{\nu} \left(
	\mathbf{h}_p^{\nu} \mathbf{P}^{\nu}
	+ \mathbf{h}_p^{L,\nu} \mathbf{P}_L^{\nu}
	+ \mathbf{h}_n^{\nu} \mathbf{N}^{\nu}
	+ \mathbf{h}_n^{L,\nu} \mathbf{N}_L^{\nu}
	\right)
	.
	\label{f}
\end{multline}

By minimizing the free energy, we get a system of site-dependent self-consistent VA equations
 to determine the values of the order parameters:
\begin{multline}
	\quad \quad 4 V L_z = H_z^L \,,\,\,
	- 4 J s^2 \mathbf{m} = \mathbf{h} \,,\,\,
	4 J s^2 \mathbf{l} = \mathbf{h}^l \,,\\
	2 t_b \mathbf{B}_0 = \mathbf{h}_b\,,\quad
	- 2 t_b \mathbf{B}_{\pi} = \mathbf{h}_b^L\,, \\
	2 t_p \mathbf{P}^{\nu} + t_{pn} \mathbf{N}^{\nu} = \mathbf{h}_p^{\nu}
	,\quad
	t_{pn} \mathbf{P}^{\nu} + 2 t_n \mathbf{N}^{\nu} = \mathbf{h}_n^{\nu}
	, \\
	- 2 t_p \mathbf{P}_L^{\nu} - t_{pn} \mathbf{N}_L^{\nu} = \mathbf{h}_p^{L,\nu}
	\, ,
	- t_{pn} \mathbf{P}_L^{\nu} - 2 t_n \mathbf{N}_L^{\nu} = \mathbf{h}_n^{L,\nu}
	.
\label{eq:orderPsys4}
\end{multline}

{\bf EF phase diagrams}
Let assume that the model cuprate described by Hamiltonian (\ref{H0}) can be found only in homogeneous phase states with long-range order determined by a single nonzero (vector) local order parameter: CO ($L_z\neq0$), AFMI (${\bf l}\neq0$), BS (${\bf B}_0\neq0$), and two types of metallic FL (${\bf P}^{\nu}\neq0 , {\bf N}^{\nu}\neq0)$ phases.
It is worth noting the specificity of the two metallic FL phases, which in our model represent a mixture of $P$- and $N$-phases due to the $PN$ ($NP$)
 contribution to the single-particle transport Hamiltonian $H_{kin}^{(1)}$, which leads to "strange"\, properties of the Fermi-type metal phases
  of cuprates with a specific coexistence of  hole and electron carriers,  characteristic of both  hole and electron  doped systems.
It is interesting that in this somewhat naive model it is possible to obtain relatively simple transcendental equations for the "critical"\, temperatures $T_{CO}$, $T_{AFMI}$, $T_{BS}$, $T_{FL}$ that determine the stability boundaries of certain homogeneous phases with one or another long-range order or corresponding second order phase transition lines\,\cite{Panov2019}.

\begin{figure*}[t]
\centering
\includegraphics[width=15 cm,angle=0]{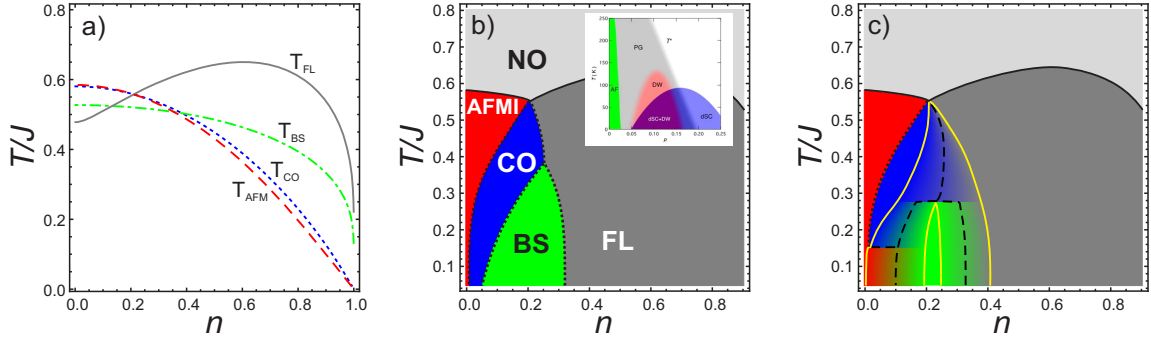}
\caption{Fig.\,1 (Color online) Model $T$\,-\,$n$ phase diagrams of the hole-doped cuprate calculated in the effective-field approximation
 ($n=p$ for the hole doping) under constant values of the Hamiltonian parameters (see text for detail); a) "critical"\, temperatures, the dashed, dotted, and dash-dotted lines indicate the boundaries of the stability region of the main homogeneous phases; b) phase diagram assuming main homogeneous phases with no allowance made for the possible coexistence of two adjacent phases;  c) phase diagram with phase separation taken into account. Black solid and dotted  curves in b) and c) point to the second and first order transition lines, respectively, dashed curves in c) point to fifty-fifty volume fraction for two adjacent phases, yellow curves in c) present the third-order phase transition lines, these limit areas with 100\% volume fraction. Inset in b) shows a typical  phase diagram observed for hole-doped cuprate\,\cite{Hamidian}. }
\label{fig1}
\end{figure*}

Making use of Exp.\,(\ref{f}) we  numerically estimated the free energies of different phases and have built a  $T$\,-\,$n$ phase diagram. In  Fig.\,\ref{fig1} we present model phase diagrams of a cuprate calculated
 given quite arbitrarily chosen parameters of the model Hamiltonian (\ref{HH})    as
$\Delta$\,=\,0.20, $V$\,=\,0.35, $t_p$\,=\,$t_n$\,=\,0.46, $t_{pn}$\,=\,0.05, $t_b$\,=\,0.65 (all in units of the exchange integral $J$). Fig.\,\ref{fig1}a shows the doping dependence of the "critical"\, temperatures $T_{CO}$, $T_{AFMI}$, $T_{BS}$, $T_{FL}$. The NO-AFMI-CO-BS-FL phase diagram is shown in Fig.\,\ref{fig1}b,  where the regions of the minimum free energy of the  phases are highlighted in different colors. Given this set of parameters, the lines of phase transitions NO-AFMI, NO-FL are lines of second-order transitions, while the lines of phase transitions AFMI-CO, CO-BS, CO-FL, BS-FL turn out to be lines of first-order phase transitions.
Comparison with the phase diagram typical of doped cuprates (see inset in Fig.\,\ref{fig1}b) shows that  the "MFA portrait"\,,  obtained under extremely simplifying assumptions,
can reproduce quite well  some principal features of the real phase diagram.
However,  a somewhat naive assumption of only homogeneous single-order-parameter phases
 may not be validated if the full multi-parameter thermodynamic field space is considered.
For instance, the free energy minimum under the assumption of a single nonzero superconducting order parameter (${\bf B}_0$) will be just a saddle point, if the nonzero charge order parameter ($L_z$)
 is also "turned on"\,, which, it would seem, should lead
  to the appearance of a homogeneous supersolid phase with the on-site CO-BS mixing.
However, despite the much more complicated Hamiltonian, the situation turns out to be absolutely similar to that implemented in the "negative-$U$"\, or hard-core boson model or a lattice model of a
superconductor with pair hopping and on-site correlation term\,\cite{Kapcia} where
 instead of forming homogeneous phases with the on-site mixing of local order parameters  the system can find it thermodynamically more convenient
to phase separate into subsystems with different volume fractions that can be readily found by adapting what is known as a Maxwell construction\,\cite{Kapcia,Arrigoni}. As it turned out as a result of the numerical implementation of the Maxwell  construction for the same parameters as above, phase separation can be realized in the region of coexistence of phases separated by a first-order phase transition line. This works for phases AFMI-FL, AFMI-BS, CO-BS, CO-FL, and BS-FL, but not for AFMI-CO. Generally speaking, in the latter case this means the possibility of the formation of a homogeneous mixed phase such as spin-charge density wave, although the effects typical for the region of the phase coexistence  will most likely be observed.
Results of the Maxwell construction for our model cuprate  presented in Fig.\,\ref{fig1}c show the significant transformation of the "naive"\, phase diagram in Fig.\,\ref{fig1}b with phase separation (PS) taken into account.
A transition between a homogeneous phase and the PS state can be
symbolically named as a "third order"\, transition with the concentration
difference as the order parameter\,\cite{Kapcia}. At this
transition a size of one domain in the PS state decreases
continuously to zero at the transition temperature.

First and second order transitions  in Fig.\,\ref{fig1}b,c are denoted by dotted and solid lines, respectively, black dashed curves point to fifty-fifty volume fraction for two adjacent phases, while yellow curves indicate "third order"\, transition, that is these delineate areas with 100\% volume fraction. It is worth noting that at "third order"\, transitions, the specific heat exhibits a finite jump as at the second order transitions\,\cite{Kapcia}.

As we see the inclusion of the PS states into consideration substantially modifies the phase diagrams of the models assuming only homogeneous phases. In the PS states the system breaks into coexisting static or dynamic domains/grains of two different phases with varying volume fraction and shape.
Hole carrier density in metallic FL phase  and in metallic domains in PS phase is $(1+p)$, however, taking into account diminishing volume fraction of metallic phase with decreasing doping we arrive at effective carrier density demonstrating the smooth crossover from $1+p$  to $p$ across optimal doping\,\cite{Pelc_2020}.
The zero resistivity transition in the phase separated state arises only when the Josephson coupling between BS domains is of the order of the thermal energy and phase locking takes place along percolating BS system. This implies a two-step superconducting transition with the formation of the isolated BS domains without phase coherence  and than by Josephson coupling with phase locking at low temperatures.

There is now considerable evidence that the tendency
toward phase separation or intrinsic electronic inhomogeneity is an universal feature of doped
cuprates(see, e.g., Ref.\,\cite{Mello} and references therein). Despite these evidences, the majority of the theoretical approaches are based on the
assumption of homogeneous phases.

It should be noted that the PS model does predict several temperatures of the "third order"\, PS transitions limiting the PS phases, that is delineating areas with 100\% volume fraction,  and the temperatures of the percolation transitions, which can manifest itself in the peculiarities of the temperature behavior for different physical quantities\,\cite{Sacksteder}.
All the phases AFMI, CO, BS are separated from the 100\% coherent metallic Fermi liquid phase by the "third order"\,  phase transition line $T^{\star}(p)$, which is  believed to be responsible for the onset of the pseudogap phenomena as a main candidate for the upper "pseudogap"\, temperature.
The PS phenomenon immediately implies an opportunity to observe as a minimum two energy pseudogaps for superconducting cuprates, related to antiferromagnetic and charge fluctuations for underdoped and overdoped compositions, respectively.
 In general, the enigmatic pseudogap phase in doped cuprates seems to be an inhomogeneous system of static and dynamic fluctuations, to be a precursor for long-range orderings, both for the CDW and dBS phases.

As we see the EF approach, realized under extremely simplifying assumptions,  is able to  reproduce essential features of the phase diagram for doped cuprates, however,  for an adequate description of real phase diagrams in the framework of the EF theory, it is necessary to take into account a number of additional vital effects.
 First of all, this concerns the real inclusion of electron-lattice polarization effects, long-range inter-site (nonlocal) correlations, and inhomogeneous potential in cuprates with nonisovalent substitution.
 As a result, we must increase the number of possible phase states, first of all, by introducing new commensurate or incommensurate spin-charge modes, or spin-charge density waves like stripes, and also take into account the screening effect of local and nonlocal correlations. The latter effect can be accurately described only with a rigorous consideration of the electron-lattice polarization effects. Experimental data point to a dramatically enhanced screening of  Coulomb interactions in cuprates under doping\,\cite{Ando,Gorkov}.
 In addition, all the "effective"\, transfer integrals $t_{p,n,pn}$ and $t_b$ will depend on the doping level through the effects of "vibronic"\, reduction.
Furthermore, our version of the effective field model assumed the use of the simplest version of the Caron-Pratt method\,\cite{Caron-Pratt} for the "real-space"\, description of one-particle transport, which seemingly leads to an overestimation of the contribution of one-particle kinetic energy.
  Specific feature of doped cuprates with nonisovalent substitution is the presence of centers of an inhomogeneous electric field, which are the nucleation centers for nanoscopic  regions of condensed charge fluctuations, providing an efficient screening of the impurity Coulomb potential.  Inhomogeneous potential will largely destroy long-range order and  lead to strong spatial fluctuations of the effective energy parameters and critical temperatures\,\cite{Pelc}.

Despite all advantages of the simple EF-MFA approach realized above, a detailed meaningful comparison with ever-expanding set of experimental data unavoidably requires the inclusion of novel effects
 about the mean field. First it concerns the effects of low-dimensionality and nonlocal quantum fluctuations.
Obviously, the effective field theory cannot provide an adequate quantitative, and in some cases, possibly even qualitative, description of low-dimensional 2D systems.
The 2D systems, in particular, the S=1 pseudo(spin) system is prone to a creation of different topological structures,  which form topologically protected inhomogeneous distributions of the eight local S=1 pseudospin order parameters\,\cite{Moskvin-JSNM-2019}. Puzzlingly, these unconventional structures can be characterized by a variety of unusual properties, in particular, filamentary superfluidity in antiphase domain walls of the CO phase and unusual skyrmions.
Main limitation of mean field theory is the neglect of correlations between spins or pseudospins i.e. effective replacement of nonlocal correlators such as $\langle{\bf \hat B}_i{\bf \hat B}_j\rangle$ by a simple product of local order parameters $\langle{\bf \hat B}_i{\bf \hat B}_j\rangle \rightarrow \langle{\bf \hat B}_i\rangle\langle{\bf \hat B}_j\rangle $. The MFA describes a long-range order of local order parameters, however, it cannot describe its precursor, that is short-range fluctuations which are of a principal importance near the critical temperatures.
One of the advantages of the EF-MFA variant used by us is the exact quantum-mechanical description of local correlations, however, the classical nature of the molecular fields leads to fundamental problems in the description of the ground state which are characteristic even of the simplest quantum antiferromagnets. Indeed,  the true ground state of the  s=1/2 antiferromagnet (given even number of spins) is a quantum superposition of all possible states with full spin S=0 and zero value of the local order parameter: $\langle {\bf s}_i\rangle$\,=\,0. The N\'eel state is just a classic "component"\, of this "hidden"\, quantum state, so-called "physical"\, ground state.
The N\'eel phases start to form at high temperatures in the nonordered phase, when thermal fluctuations and fluctuating non-uniform fields  destroy the quantum states, while the N\'eel-type domains become more and more extended and stable with decreasing the temperature,  leaving no real chance of the formation of a true quantum ground state in the low-temperature limit.
The contribution of purely quantum states is manifested in a significant decrease in the value of the local order parameter in the N\'eel "portrait"\, as compared with the nominally maximum value of $s$. It should be noted that all the   phases with long-range order we address above, AFMI, CO, BS, are N\'eel like, that is these are characterized by a nonzero local order parameters.
As in quantum magnets, the existence of the "MFA-hidden"\, quantum state in HTSC cuprates leads to a significant suppression of the magnitude of the local order parameters for CDW and superconducting (BS) phases\,\cite{Bozovic,Sushkov}. Thus, the EF-MFA phase diagram we are considering "hides"\, the existence of a true quantum ground state, a "quantum background"\,, such as the Anderson's RVB (resonating valence bond)  phase\,\cite{RVB}, formed by a system of EH dimers\,\cite{Moskvin-Panov-PSS-2019}.

In summary, we have presented an unified non-BCS approach to the description of the variety of the local order parameters and the single local order parameter phases in high-$T_c$ cuprates.
Instead of conventional quasiparticle $\bf k$-momentum description we made use of a real space on-site "unparticle"\, S=1 pseudospin formalism to describe the charge triplets and introduce an  effective spin-pseudospin Hamiltonian which takes into account main on-site and inter-site interactions. We performed the analysis of the ground state and $T$\,-\,$n$ phase diagrams of the model Hamiltonian by means of a site-dependent variational approach in the grand canonical ensemble within effective field approximation typical for spin-magnetic systems. Within two-sublattice approximation and $nn$-couplings we arrived at several MFA, or N\'{e}el-like phases in CuO$_2$ planes with a single nonzero local order parameter: antiferromagnetic insulator, charge order, glueless $d$-wave Bose superfluid phase, and unusual metallic phase. However, the global minimum of free energy is  realized for inhomogeneous phase separated states which emerge below temperature $T^{\star}(p)$, which is believed to be responsible for the onset of the pseudogap phenomenon. With a certain choice of the Hamiltonian parameters the model EF phase diagram can quite reasonably reproduce the experimental phase diagrams.

Our thanks to V.Yu. Irkhin for fruitful discussions.
This research was funded by Act 211 Government of the Russian Federation, agreement No 02.A03.21.0006 and by the Ministry of Education and Science, project No FEUZ-2020-0054.

\end{document}